# Galaxy Mass Distributions for Some Extreme Orbital-Speed Profiles

Kenneth F. Nicholson (Caltech Alumni)
email: knchlsn@alumni.caltech.edu

## Abstract

Extreme speed profiles, that are impossible to analyze correctly using the usual  methods, are shown to be easily handled using the methods developed previously by Nicholson (astro-ph / 0006330). These profiles are constant orbital speeds with sharp and smooth ramps at the origin, and constant angular velocity with speeds increasing linearly to the galaxy rim.  Also an example using measured data for a real galaxy is shown for NGC 3198.  A new format is presented to show results in dimensionless form, that includes a way to show SMD results so that they are easily compared. Based on repeated trials for accuracy, the rod representing the mass of a fundamental segment is now placed at 0.575 *dr  at the center ring of a galaxy.

## Introduction

The orbital speeds of many thin-disk galaxies, after rising at the center, are nearly constant to large radii. Analysts have had to invent some unusual methods to find a reasonable mass distribution that could cause these speeds. Most prominent of these methods is the dark-matter halo, a distribution of spherical halos with masses that adjust the speeds where needed. These halos cannot be detected in any physical way and their existence is only implied by their need to make the mathematics work using "Newtonian mechanics."  Newton's law has been checked by many applications, but it is not applied correctly there. A new field of astro-physics has been created in an attempt to find the "dark matter" that fills these halos, so far with no success.

 In spite of the great popularity of these halos, I feel they probably don't exist. The emperor has no clothes.  Nearly all the matter in a galaxy lies within the detectable (by any means) thin-disk envelope. Light has no effect on dynamic results and it is a fair bet that most of the galaxies in the universe are small and "dark" in the traditional sense. But this is not really mysterious dark matter.

Another scheme invented in an attempt to avoid the halos is the Modified Newtonian method, where the gravitational constant is adjusted with distance. Unfortunately most of these schemes are based on the assumption that the mass of all matter inside the radius of a measured speed is determined by that speed, and the effects of matter outside that radius on that inside, are zero. This assumption is false except for spheres, and Newton's method must be applied by integrating the gravitational effects of ALL parts of the galaxy on a "test mass" at a given radius, to find the speed there.

## 1.  Method

Using the equations as developed by Nicholson (2000), a fixed thickness distribution is assumed for the galaxy, a starting value is used for density, and the speed profile computed.  The speed errors (measured minus computed) at each of the ring outer radii are then used to correct the densities of the rings and the process repeated until all speed errors are less than vmax *1E-6.

. If no better thickness distribution is known, the Milky Way thickness shape is used. Zero thickness cannot be used because that causes low values for both SMD and total mass. Also zero thickness leads to division by zero in the computing. The effects of thickness have been previously shown by Nicholson.

The digital form of the equation for orbital speed at the test mass radius rt is:

$$v^2 = (pytks)^2 \cdot rt \cdot \sum_{1}^{Nr} 2 \sum_{1}^{180} \frac{G\ ddm}{\sqrt{c^2 + (h/2)^2}} \cdot \frac{(rt - rr\cos(th))}{c^2}$$

where  $c^2 = (rr\sin(th))^2 + (rt - rr\cos(th))^2$,  pc^2

  ddm = mass of the fundamental segment, rho * r * dth * dr * h,  msuns
  dth  = 1 degree
  G    = gravitational constant, 4.498E-15 pc^3/(msuns/yr^2)
  h    = galaxy thickness at radius r, pcs
  Nr   = number of rings ( 30 for all cases here except Milky Way has 20)
  pytks = 9.778E5  (kms/sec)/(pys/yr)
  r    = radius to centerline of ring, pcs
  rho  = density, msuns/pc^3
  rm   = radius to outer edge of ring, pcs
  rr   = radius to rod used to represent fundamental segment mass, pcs
       = rm -dr / 2 * (rm - 0.575 * dr) / (rm - dr / 2)
  rt   = radius to test mass, pcs
  v    = orbital speed of test mass, kms/sec

## 2.  Results

All results are given in dimensionless form so that galaxies of different sizes are directly comparable for dynamic effects. A new parameter, a dimensionless form of SMD, allows easy comparison for different galaxies.  Data are given on each plot allowing reduction to dimensioned data if desired. Dimensionless forms are:

  md    = m / mtot
  rd    = r / rmax
  rhod  = rho / rhoav
  rhoav = mtot / voltot, msuns / pc^3
  vd    = v / vkrim
  vkrim = Kepler  result at the rim, sqr (G * mtot / rmax) * pytks, kms/sec
  rSMDd = rd * SMD / SMDav
  SMDav = mtot / (pi * rmax^2),  msuns / pc^2

The Milky Way shape was obtained from a picture in a Bok article (1981), and seems to give reasonable results for densities.

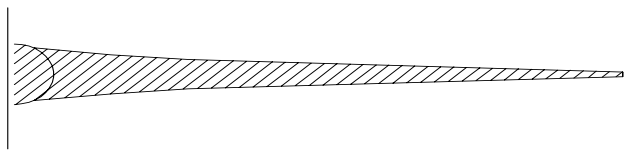

Milky Way shape

hd = 0.1288*sqr(1 - (rd/0.0644)^2) ,  for rd < 0.0322 ,
then   hd = 0.11158*exp ( - 2.315*(rd - 0.0322)) , for rd < 0.2857 ,

    then    hd = 0.0620 - 0.0727 * (rd - 0.2857)
  where    hd = h / rmax

All examples use the Milky Way galaxy shape, except for that having constant angular velocity, which uses constant thickness. When the thickness is reasonably correct, the mass distribution (md vs rd) and SMD depend slightly on the thickness selection if hd is "small" ( ie. thin disk), and density has the largest errors. Fairly large errors and computing problems result if the thickness is assumed to be zero. Density of course is inversely proportional to the galaxy thickness assumed, so the errors there can be of order 25%. Better data are needed for thickness distributions to improve density output. As experience accumulates, the thickness distributions for each galaxy considered will become more accurate.

The constant-speed example of figure 1, using a sharp transition, results in a spike in the rSMD curve at the transition. The slope of that curve near the origin is proportional to SMD at the center. All examples become asymptotic to the Kepler result beyond the rim as they should. No other methods have so far shown results beyond the rim, although some could be extended to that region. Because the speed errors are so small, the "measured" and computed speeds are essentially identical. These figures can be enlarged easily in PDF. For this trial rmax = 12000 pcs.

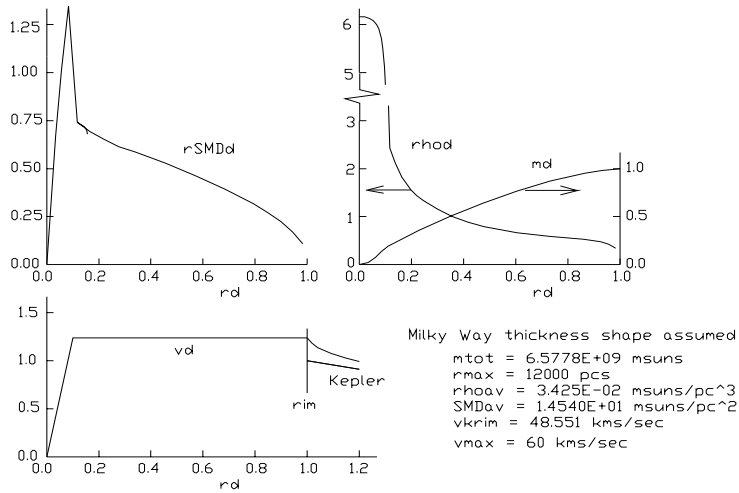

Figure 1. Constant orbital speeds with sharp transition

This spike is eased using a smooth transition as in figure 2. Note rmax changes from 12000 to 18000.

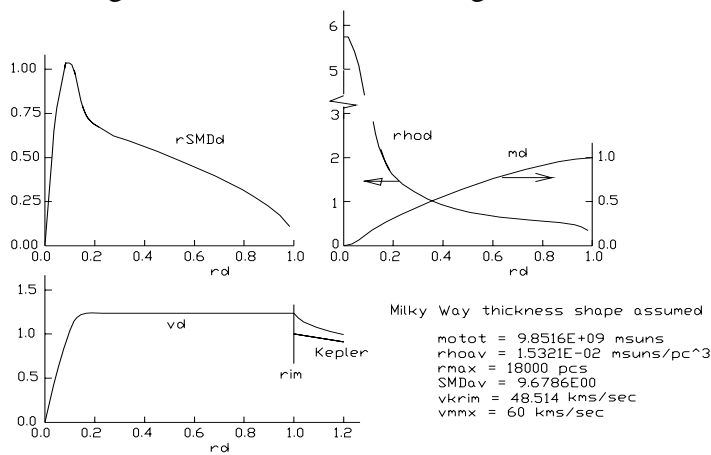

Figure 2. Constant orbital speeds with smooth transition

NGC 3198, has almost constant rotational speeds from 6 to 30 kpcs and the results (figure 3) are similar to those of figures 1 and 2, except for the bumpiness caused by the variations in measured speeds. The speeds used are from figure 10-2 of Binney and Tremaine. Rmax is now 30000.

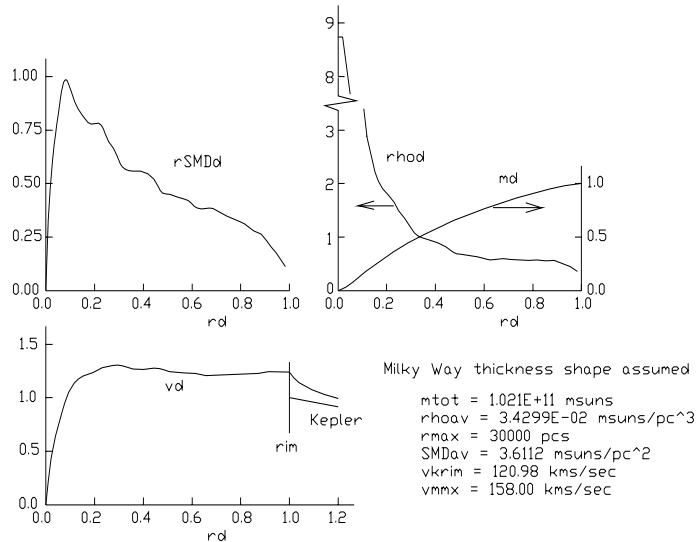

Figure 3. NGC3198 dimensionless parameters

When plotted together in figure 4, the results of figures 1, 2, 3 show the similarities and differences.

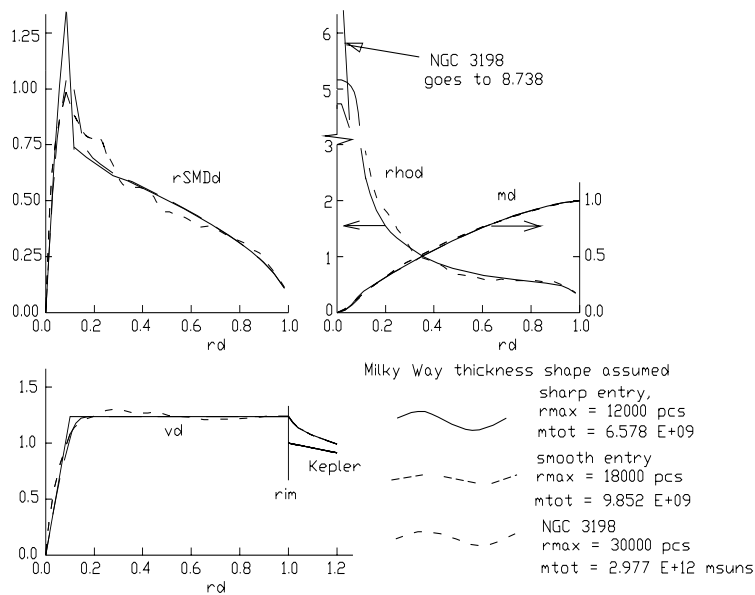

Figure 4. Three dynamically similar galaxies

This next "galaxy" does not seem to occur naturally, but there is no dynamic reason for it not to exist. It is a constant-thickness disk with constant angular velocity. That is, the speeds are proportional to radius, and no shear can occur between rings. It is essentially a solid flywheel. There appears to be no known analytic solution for the mass distribution, but using Nicholson's method, the SMD and density are easily obtained (figure 5). The curves show the SMD and density are constant near the center, falling off to small values near the rim. A result using constant density and solving for thickness would have a slightly different rSMD curve.

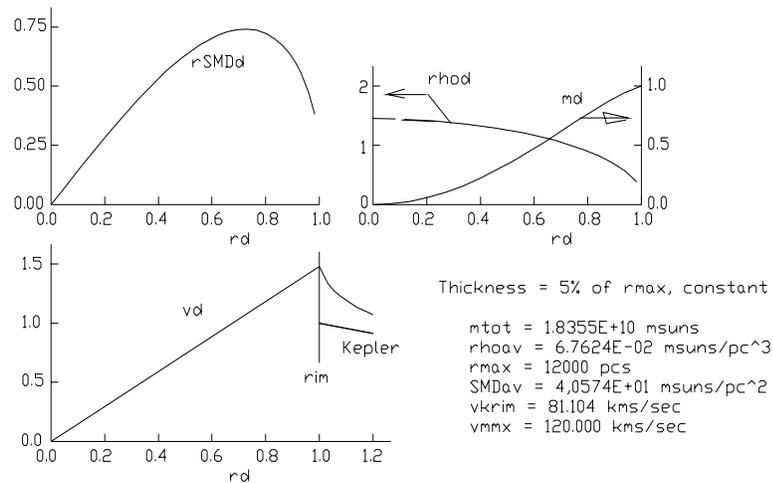

Figure 5. Constant angular velocity

## 3. Conclusions

Newton's law needs no correction to find galaxy mass distributions from orbital speeds.

Given reasonable estimates of galaxy dimensions, including thickness, good values for the mass, SMD, and density distributions are easily found from the orbital speed profiles.

There is no need for spherical dark-matter halos to find a mass distribution to match the orbital apeed profiles for thin disks, and these halos probably don't exist.

**References**

Binney, J. and Tremaine, S. 1987, Galactic Dynamics (book, Princeton University Press)
Bok, R.J., The Milky Way Galaxy, Scientific American, March 1981
Nicholson, K. F., astro-ph/0006330